\documentclass[aps,prc,twocolumn,times,graphicx,tighten,showpacs]{revtex4}
\usepackage[dvips]{graphicx}
\usepackage[dvips]{color}

\newcommand{\be}{\begin{equation}}
\newcommand{\ee}{\end{equation}}
\newcommand{\ba}{\begin{eqnarray}}
\newcommand{\ea}{\end{eqnarray}}

\newcommand{\nr}{{\bf      r}}

\begin{document}

\title{Superscaling predictions for NC and CC 
quasi-elastic neutrino-nucleus scattering} 

\author{
J.E. Amaro$^a$,
M.B. Barbaro$^b$,
J.A. Caballero$^c$,
T.W. Donnelly$^d$,
}

\affiliation{$^a$Departamento de F\'{\i}sica At\'omica, Molecular y Nuclear,
Universidad de Granada, Granada 18071, Spain}

\affiliation{$^b$Dipartimento di Fisica Teorica, Universit\`a di Torino and
  INFN, Sezione di Torino, Via P. Giuria 1, 10125 Torino, Italy}

\affiliation{$^c$Departamento de F\'{\i}sica At\'omica, Molecular y Nuclear,
Universidad de Sevilla, Apdo.1065, 41080 Sevilla, Spain}

\affiliation{$^d$Center for Theoretical Physics, Laboratory for Nuclear
  Science and Department of Physics, Massachusetts Institute of Technology,
  Cambridge, MA 02139, USA}

\begin{abstract}
Quasielastic double differential neutrino cross sections can be
obtained in a phenomenological model based on the superscaling
behavior of electron scattering data. In this talk the superscaling
approach (SuSA) is reviewed and its validity is tested in a
relativistic shell model. Results including meson exchange currents 
for the kinematics of the MiniBoone experiment are presented.
\end{abstract}


\maketitle

Analysis of inclusive $(e,e')$ data have demonstrated
that at energy transfers below the quasielastic (QE) peak superscaling is
fulfilled rather well \cite{Day90}---\cite{Don99b}.  The general
procedure consist on dividing the experimental $(e,e')$ cross section
by an appropriate single-nucleon cross section to obtain the
experimental scaling function $f(\psi)$, which is then plotted as a
function of the scaling variable $\psi$ for several kinematics and for several nuclei. 
 If the results do not depend
on the momentum transfer $q$, we say that scaling of the first kind
occurs. If there is not dependence on the nuclear species, one has
scaling of the second kind. The simultaneous occurrence of scaling of
both kinds is called superscaling.
The Super-Scaling approach (SuSA) is based on the assumed universality
of the scaling function for electromagnetic and weak
interactions \cite{Ama05}.

\begin{figure}
\includegraphics[scale=0.4,  bb= 150 360 400 790]{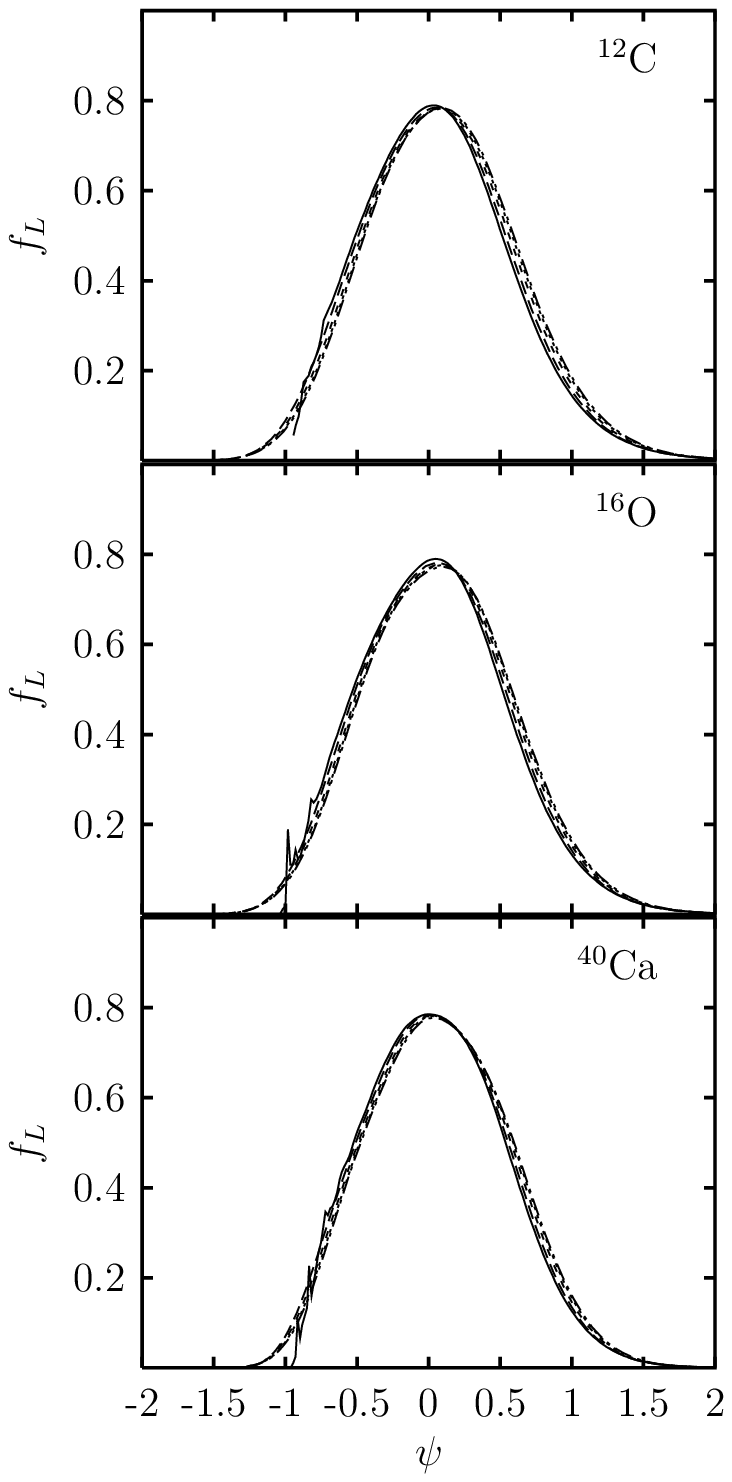}%
\includegraphics[scale=0.4,  bb= 180 310 400 790]{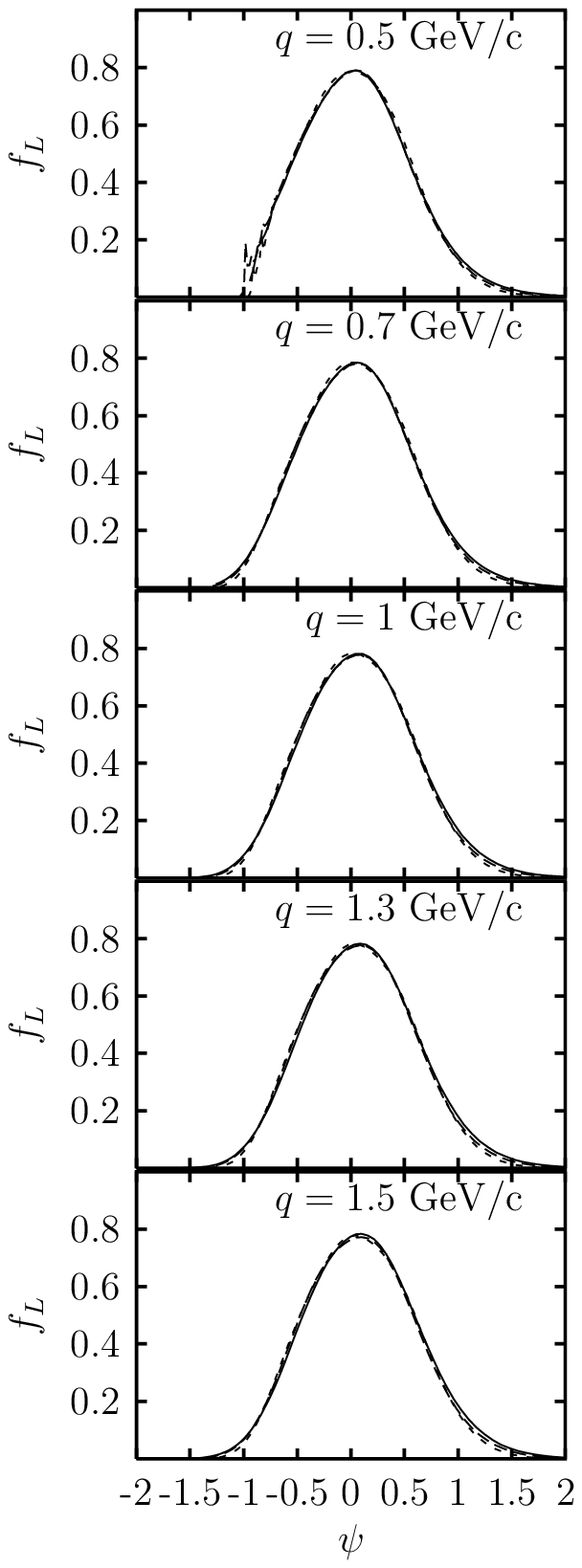}
\caption{\label{scaling-sr}
Scaling properties of the SR shell model.
Left:
scaling of the first kind.
Curves for 
$q=0.5,0.7,1,1.3,1.5$ GeV 
collapse into one.
Right: scaling of the second kind.
Curves for 
$^{12}$C, $^{16}$O and $^{40}$Ca 
collapse into one.
}
\end{figure}

\begin{figure}
\includegraphics[scale=0.6,  bb= 70 360 490 790]{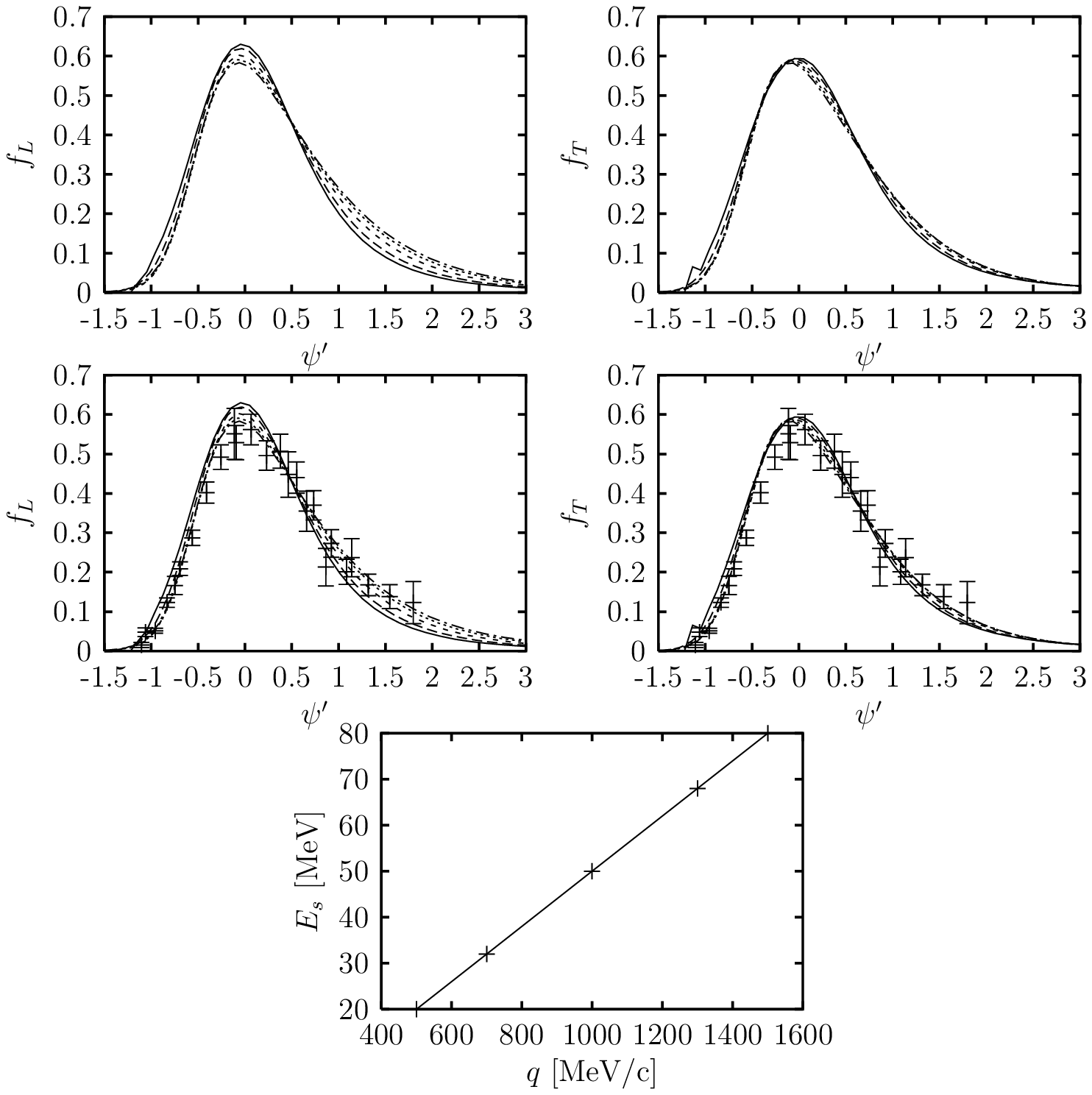}
\includegraphics[scale=0.4,  bb= 40 90 490 780]{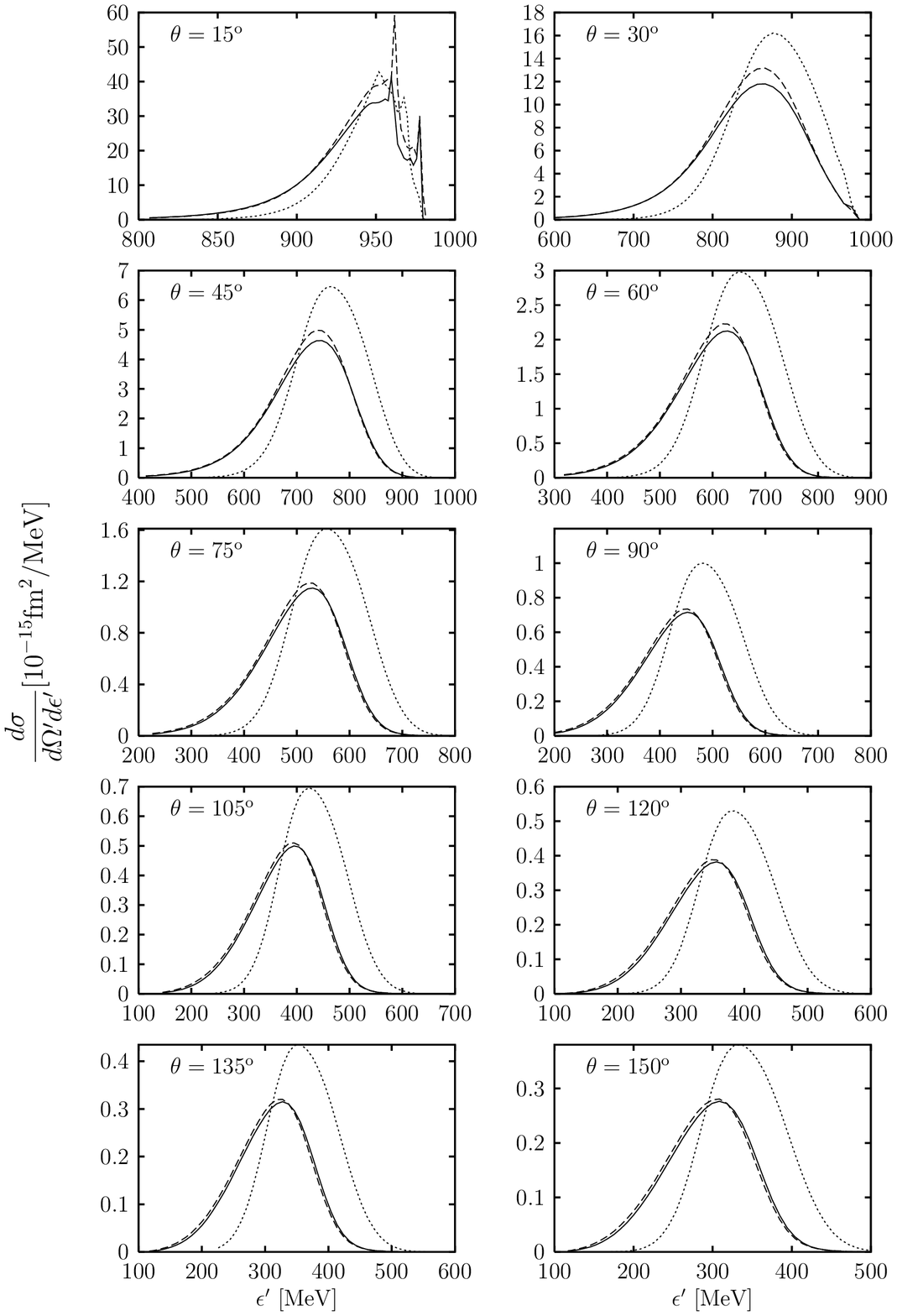}
\caption{\label{deb-scaling}
Left: Scaling 
of 1st kind with
DEB+D potential compared with experimental data.
$q=0.5$, 0.7
1.0, 1.3 
and 1.5 GEV/c.
Right:
Test of SuSA in the SR shell model for the $^{12}$C$(\nu_\mu,\mu^-)$
reaction with neutrino energy $\epsilon=1$ GeV.
Dotted: Woods-Saxon potential.
Solid: DEB+D potential.
Dashed: SuSA reconstruction from the computed $(e,e')$ scaling function.
}
\end{figure}

The superscaling property is exact in the relativistic Fermi gas
model (RFG) by construction, and it has been tested in more realistic
models of the $(e,e')$ reaction \cite{Ama05c}--\cite{Ama07}. A study of
superscaling in the semirelativistic (SR) continuum shell model with
Woods-Saxon (WS) mean potential is summarized in Fig. 1. 
There we show the longitudinal scaling function defined
as $f_{L}=R_{L}/G_{L}$, where $R_L$ is the longitudinal response
function and $G_L$ is a single-nucleon factor.  When $f_L(\psi)$ is
plotted for various values of the momentum transfer and for several
closed-shell nuclei, all the curves approximately collapse into
one. Small violations of scaling are seen at low values of $\psi$,
coming from the low-energy potential resonances for $q=0.5$ GeV/c,
which disappear for higher $q$ values.

While the SR shell model superscales, it does not reproduce the
experimental data of the phenomenological scaling function extracted
from the longitudinal QE electron scattering response. The WS
potential used to describe the final-state interaction (FSI) of the
ejected proton does not incorporate the appropriate reactions
mechanisms. A further improvement of the FSI consist in using
the Dirac-Equation based potential plus Darwin term (DEB+D).  The DEB
potential is obtained from the Dirac equation
for the upper component
$\psi_{up}(\nr)=K(r,E)\phi(\nr)$, where the Darwin term $K(r,E)$ is
chosen in such a way that the function $\phi(r)$ satisfies a
Schr\"odinger-like equation. The electromagnetic L and T scaling
functions within this model are shown in Fig. 2. We use the same
relativistic Hartree potential as in the relativistic mean field model
of \cite{Ama05c}, and  the scaling variable $\psi'$ includes a
$q$-dependent energy shift $E_s(q)$. Although the scaling is not
perfect, it is remarkable that our results give essentially the same
scaling function for a wide range of $q$ values, reproducing well the
phenomenological data.

\begin{figure}
\includegraphics[scale=0.5,bb=70 400 550 750,clip]{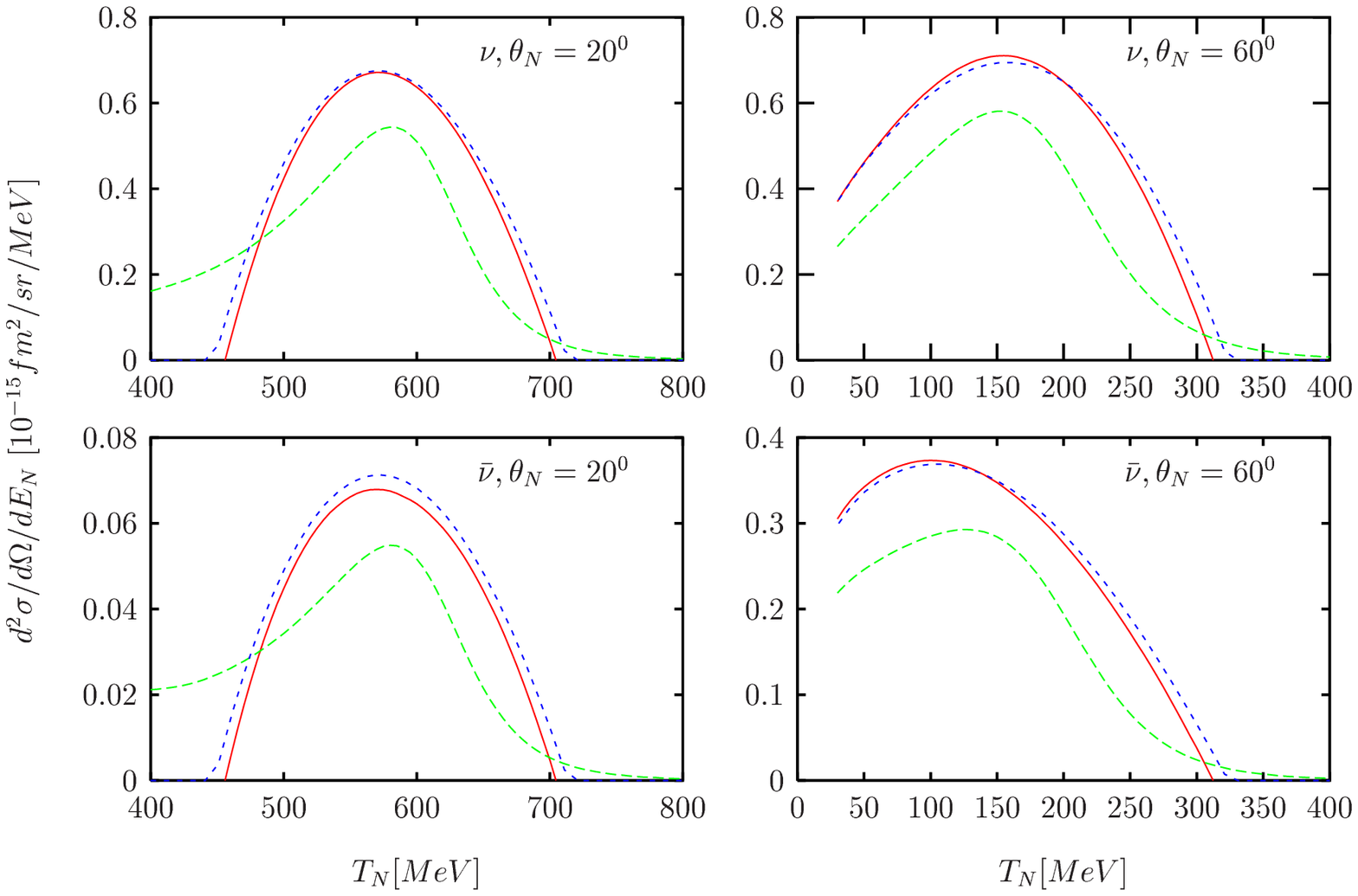}
\includegraphics[scale=0.4,bb=30 300 580 750,clip]{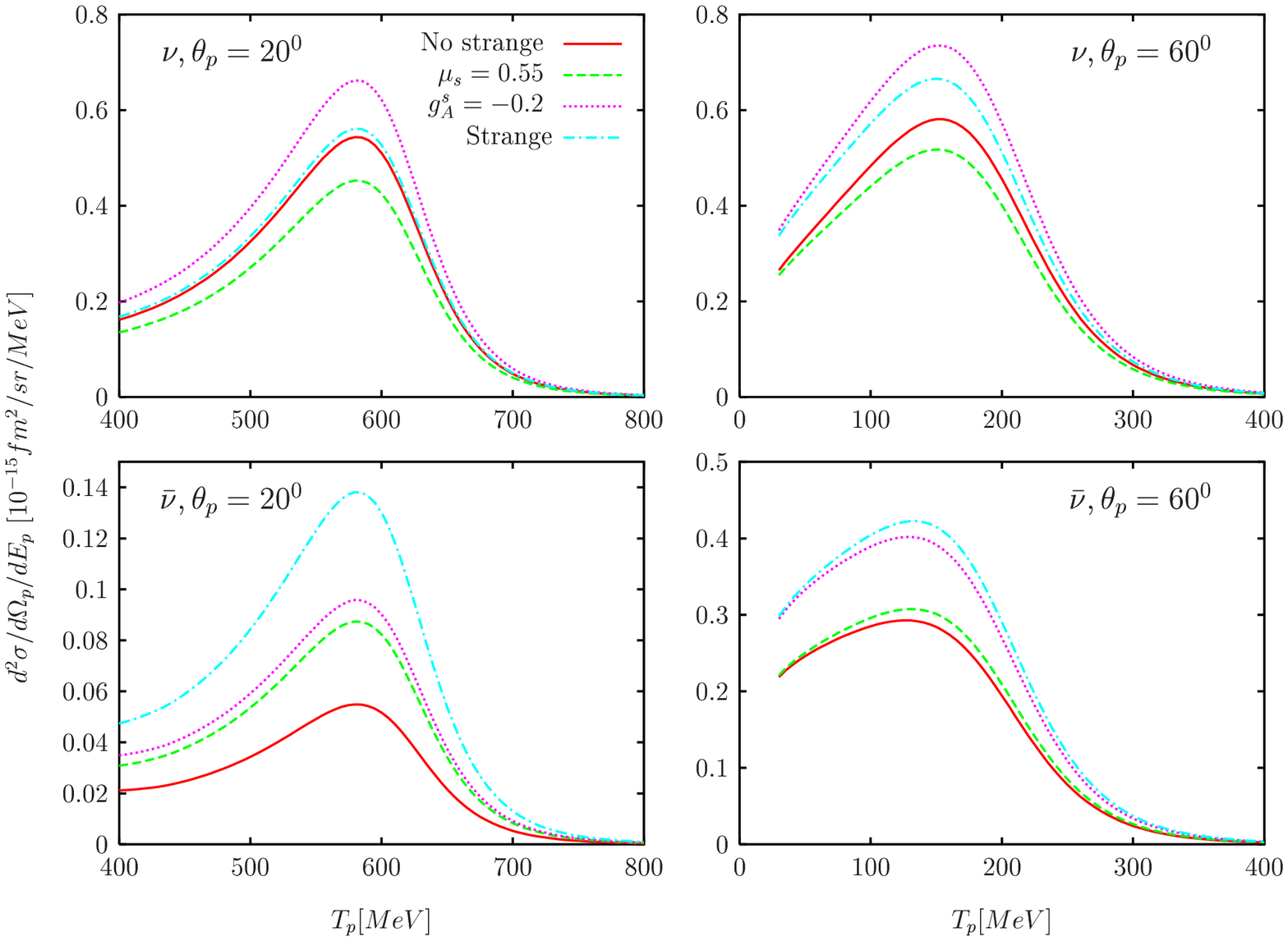}

\caption{Neutral current proton knock-out cross section from
  $^{12}$C. Left: RFG (solid lines), RFG with factorization (dotted lines), 
and SuSA (dashed lines). Right: study of nucleon
  strangeness effect.}
\end{figure}

\begin{figure}
\includegraphics[scale=0.4,bb=130 470 490 690,clip]{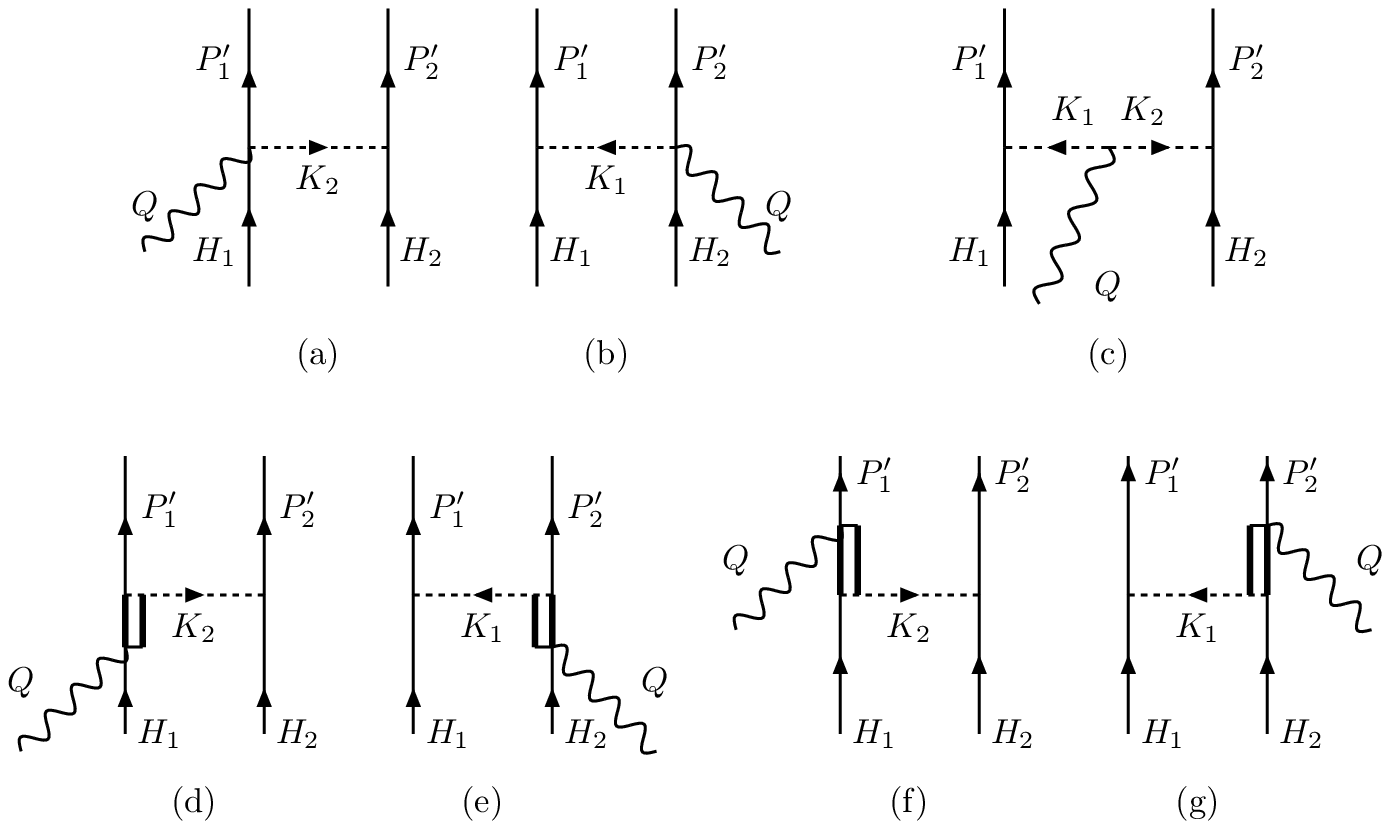}
\caption{Diagrams contributing to the MEC}
\end{figure}

\begin{figure}
\scalebox{0.7}{
\parbox{10cm}{
\includegraphics[scale=0.35]{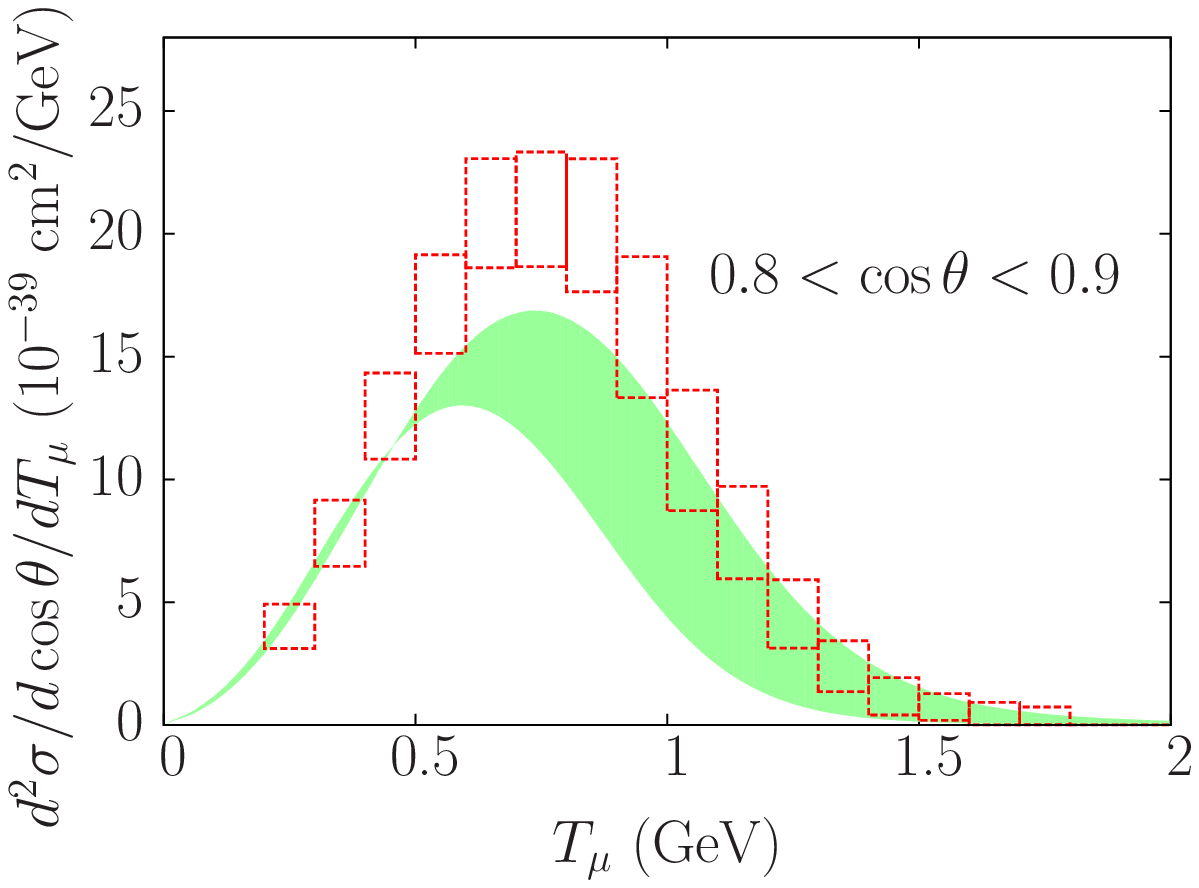}%
\includegraphics[scale=0.35]{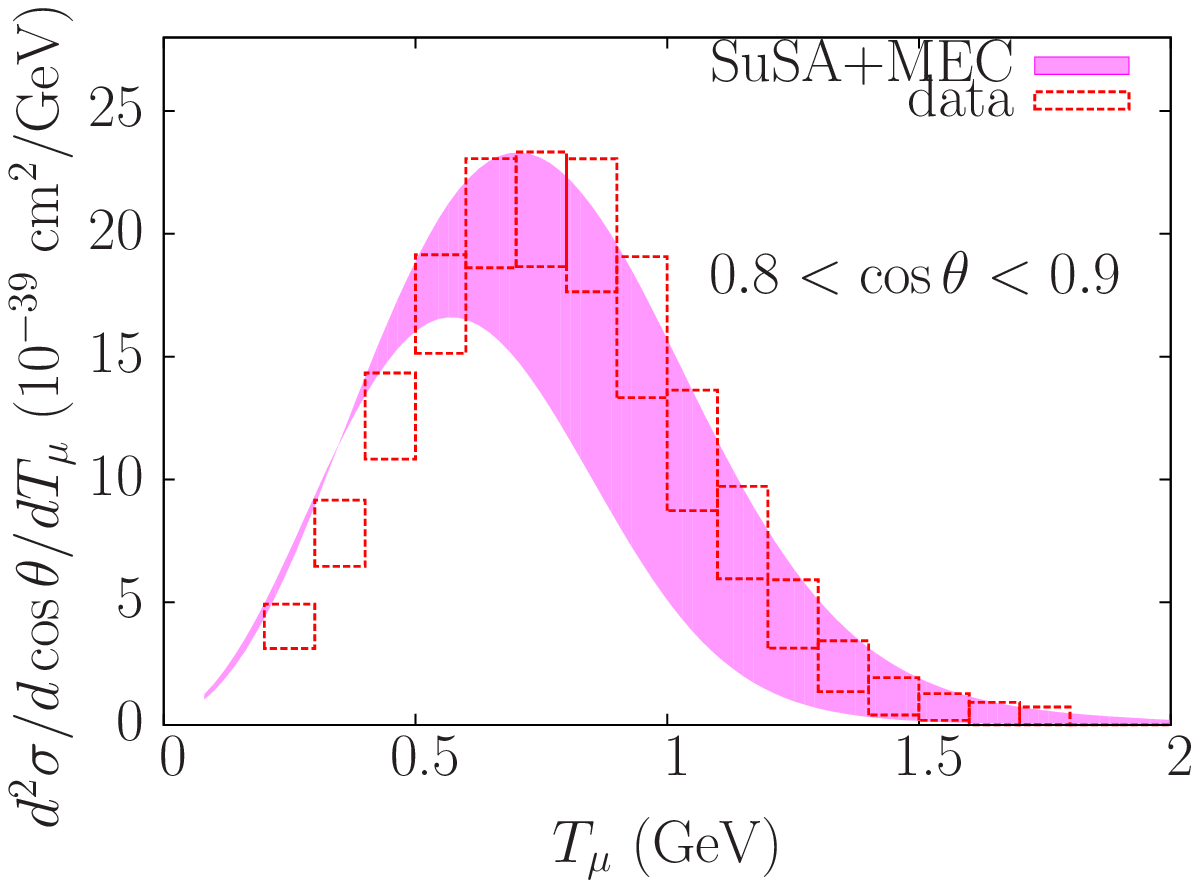}%
\\
\includegraphics[scale=0.35]{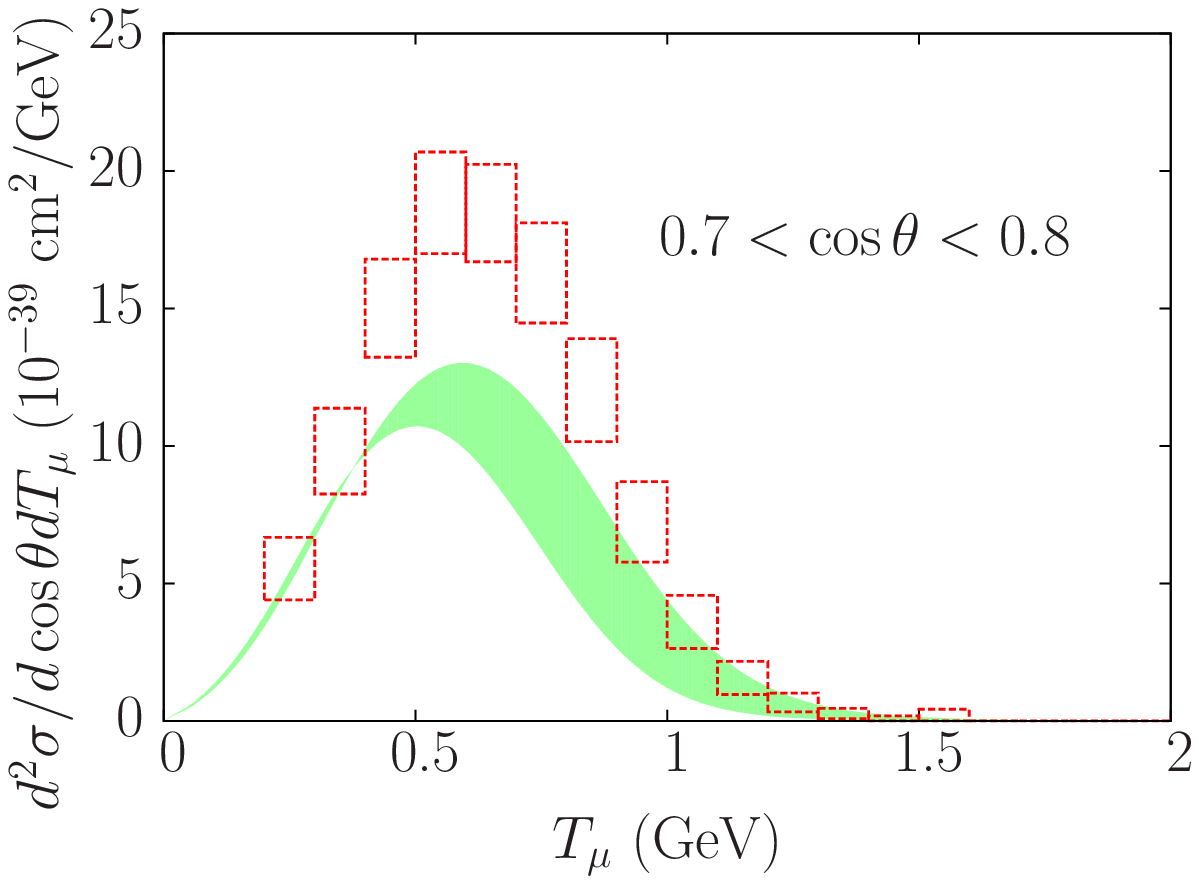}%
\includegraphics[scale=0.35]{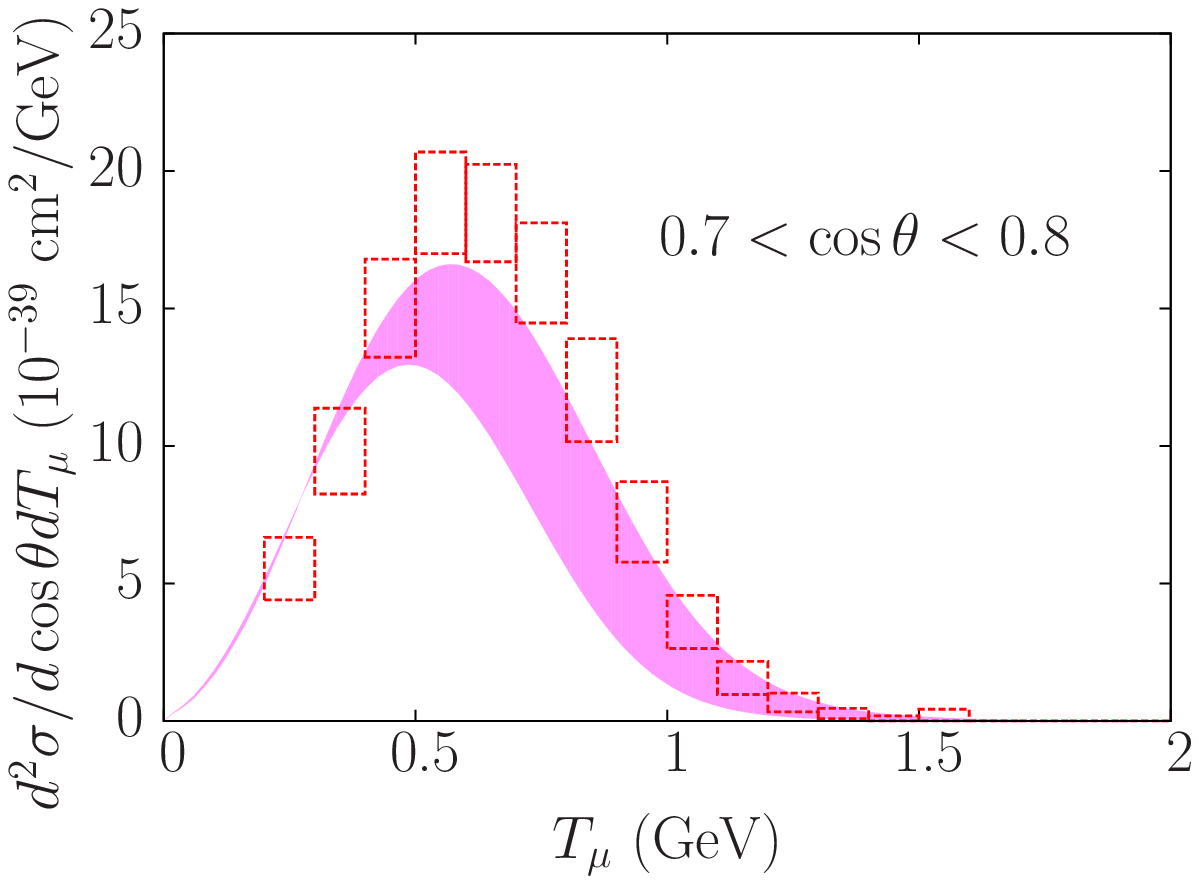}%
\\
\includegraphics[scale=0.35]{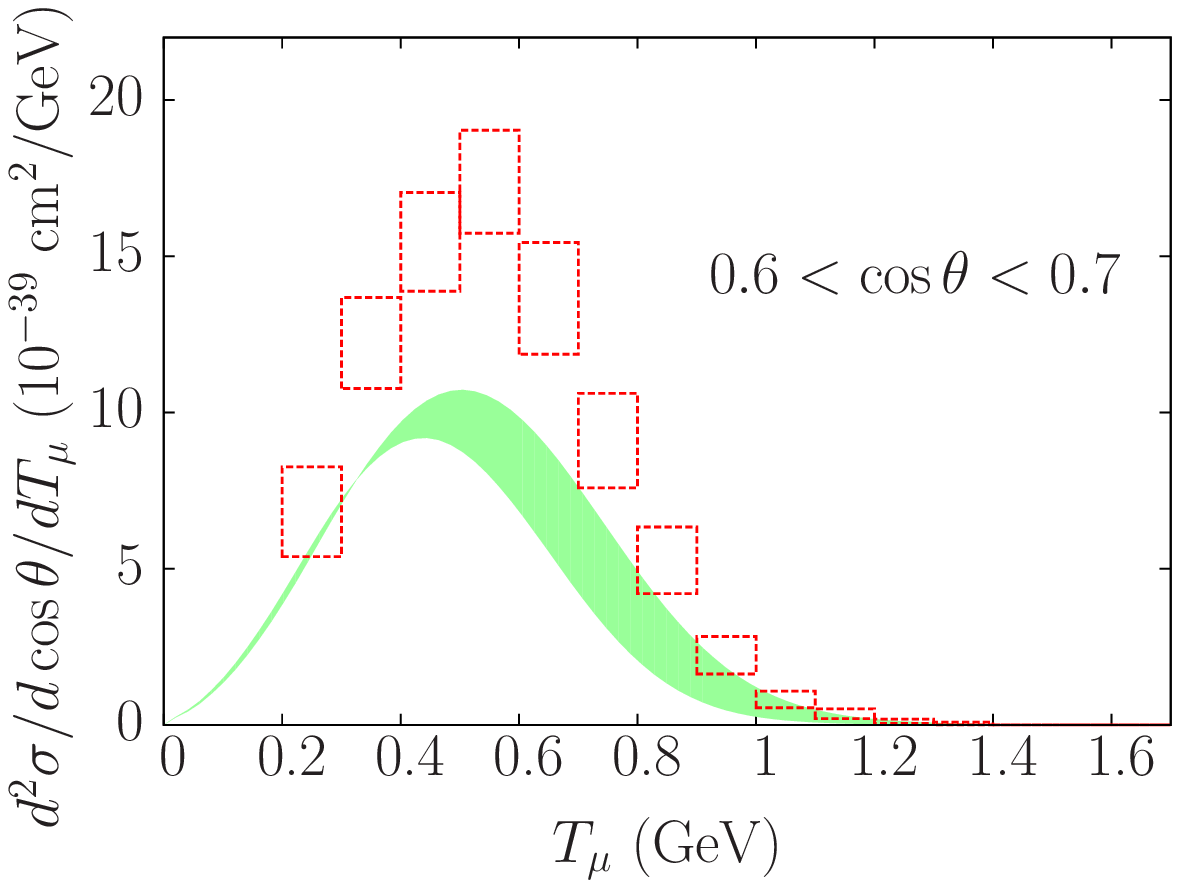}%
\includegraphics[scale=0.35]{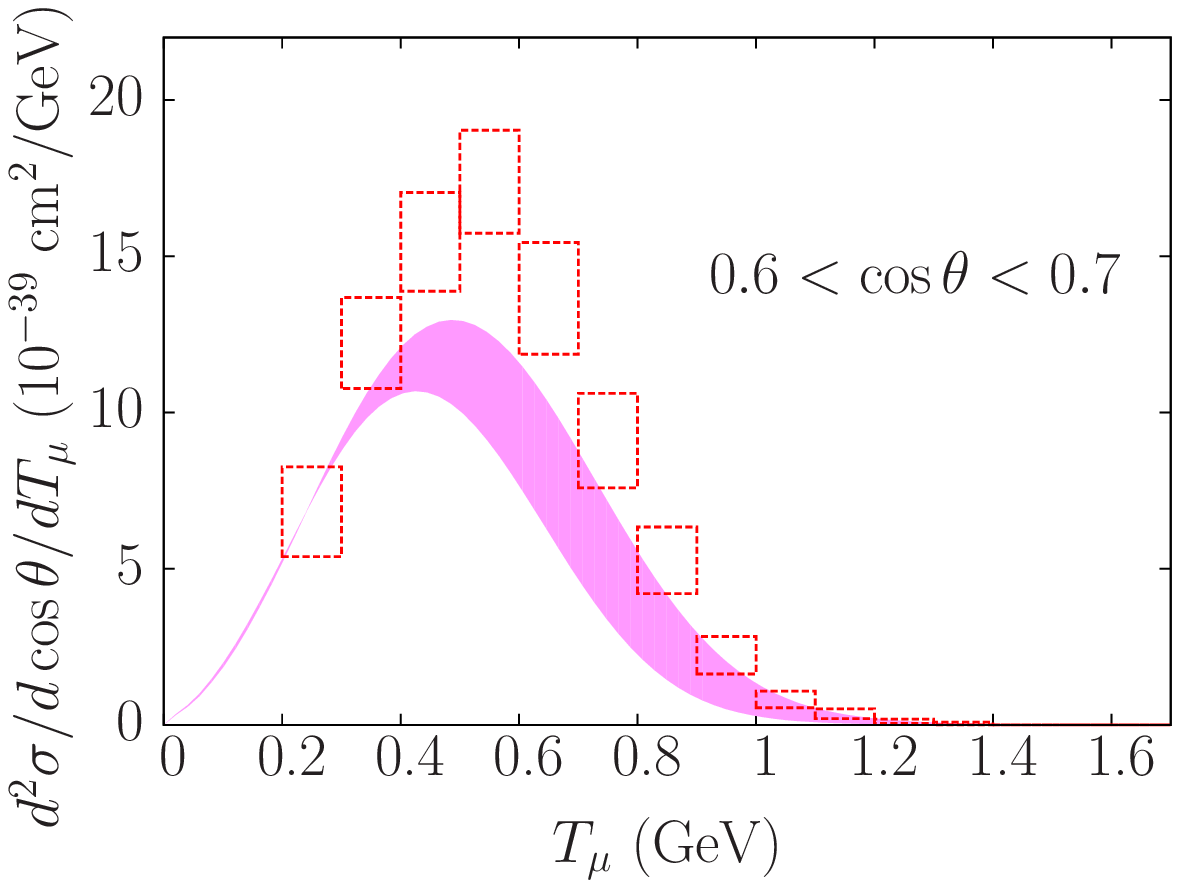}%
\\
\includegraphics[scale=0.35]{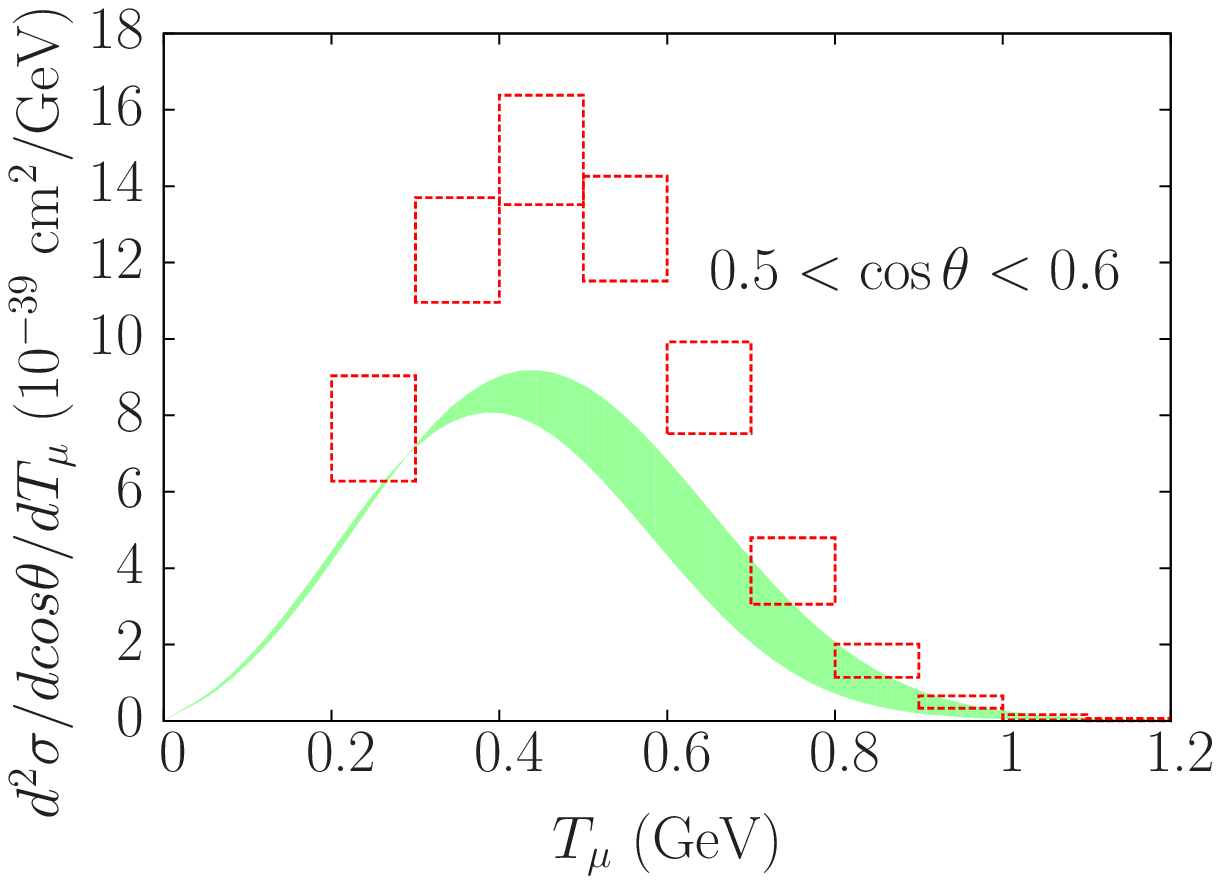}%
\includegraphics[scale=0.35]{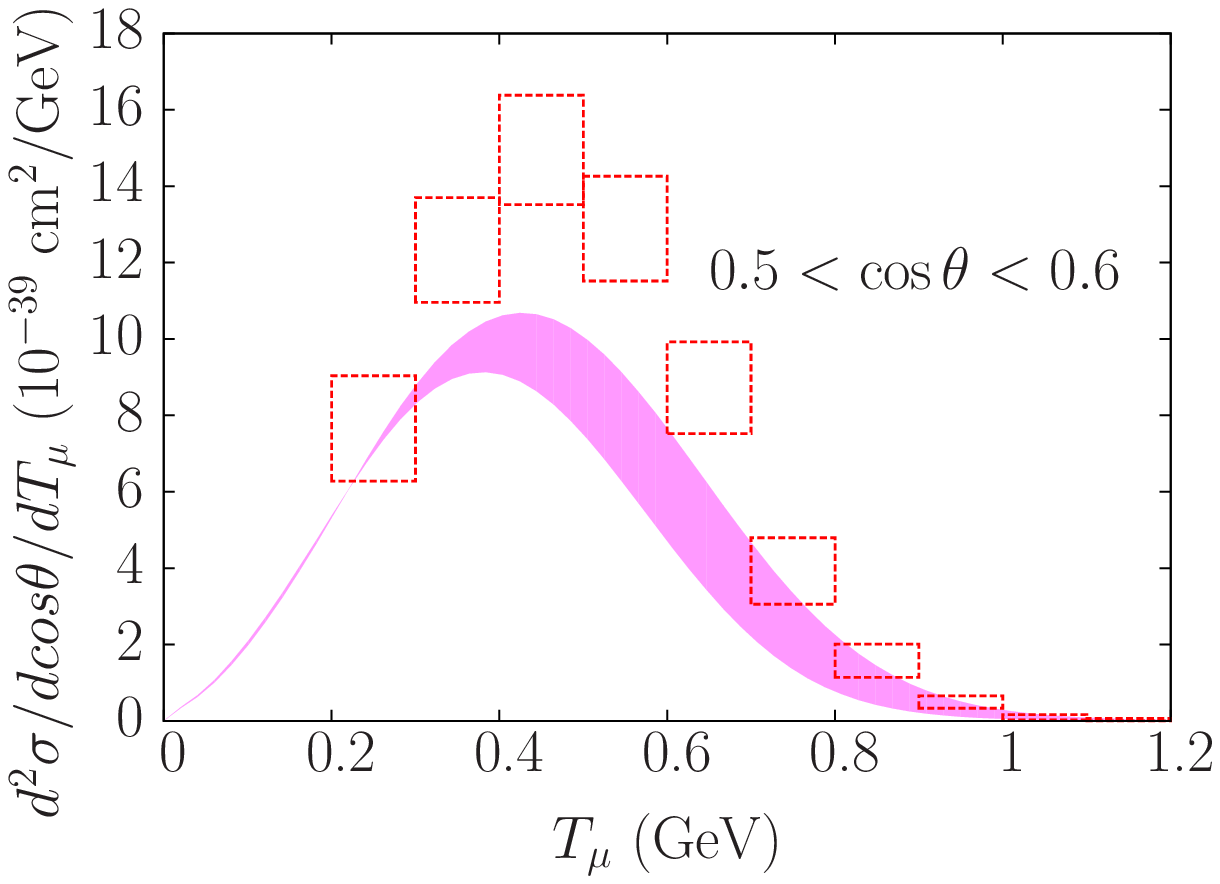}%
\\
\includegraphics[scale=0.35]{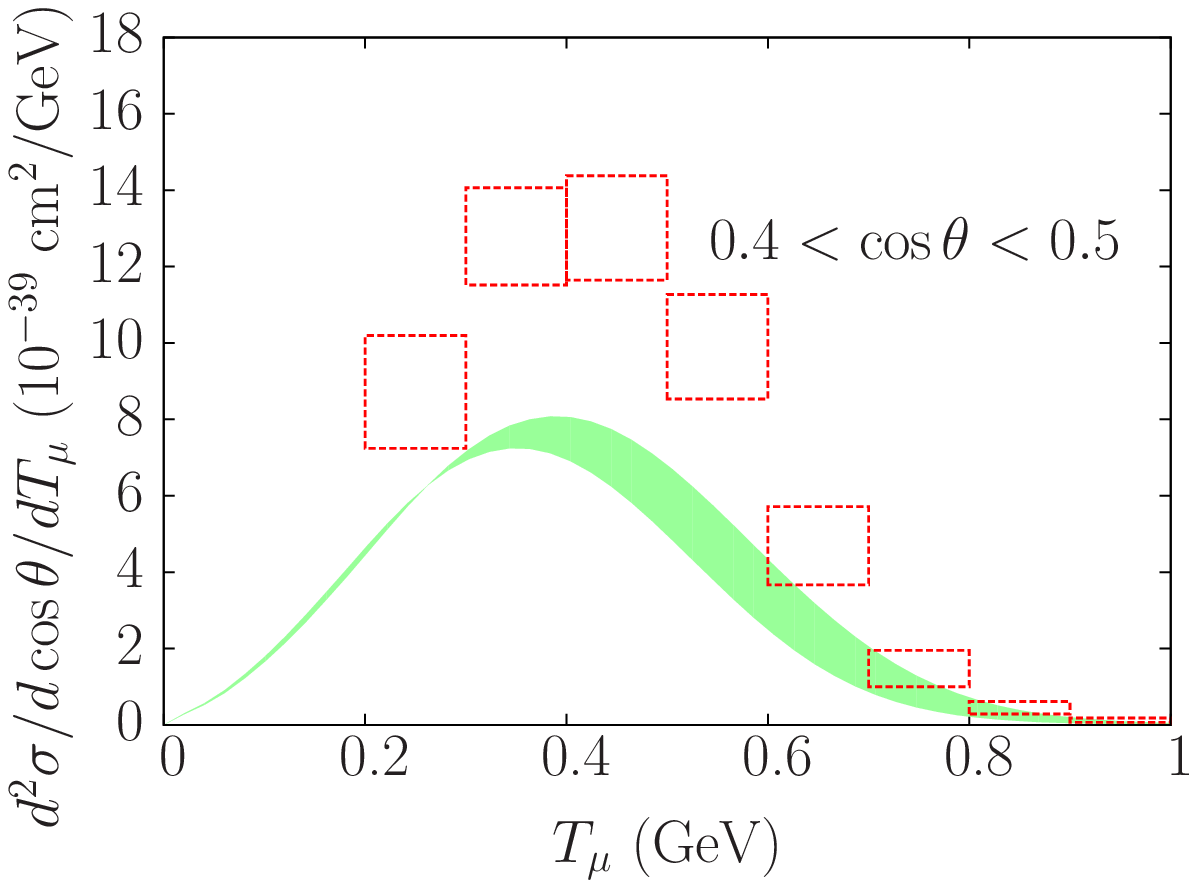}%
\includegraphics[scale=0.35]{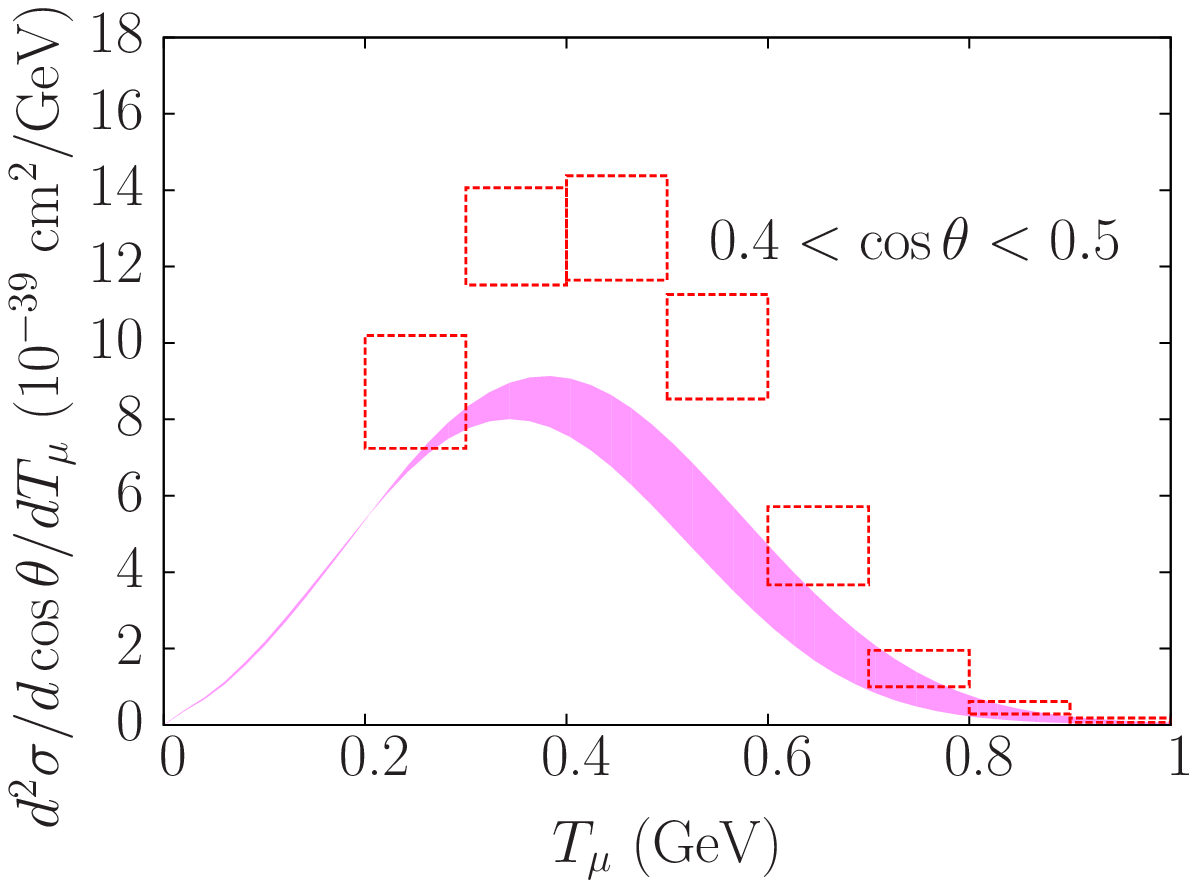}%
\\
\includegraphics[scale=0.35]{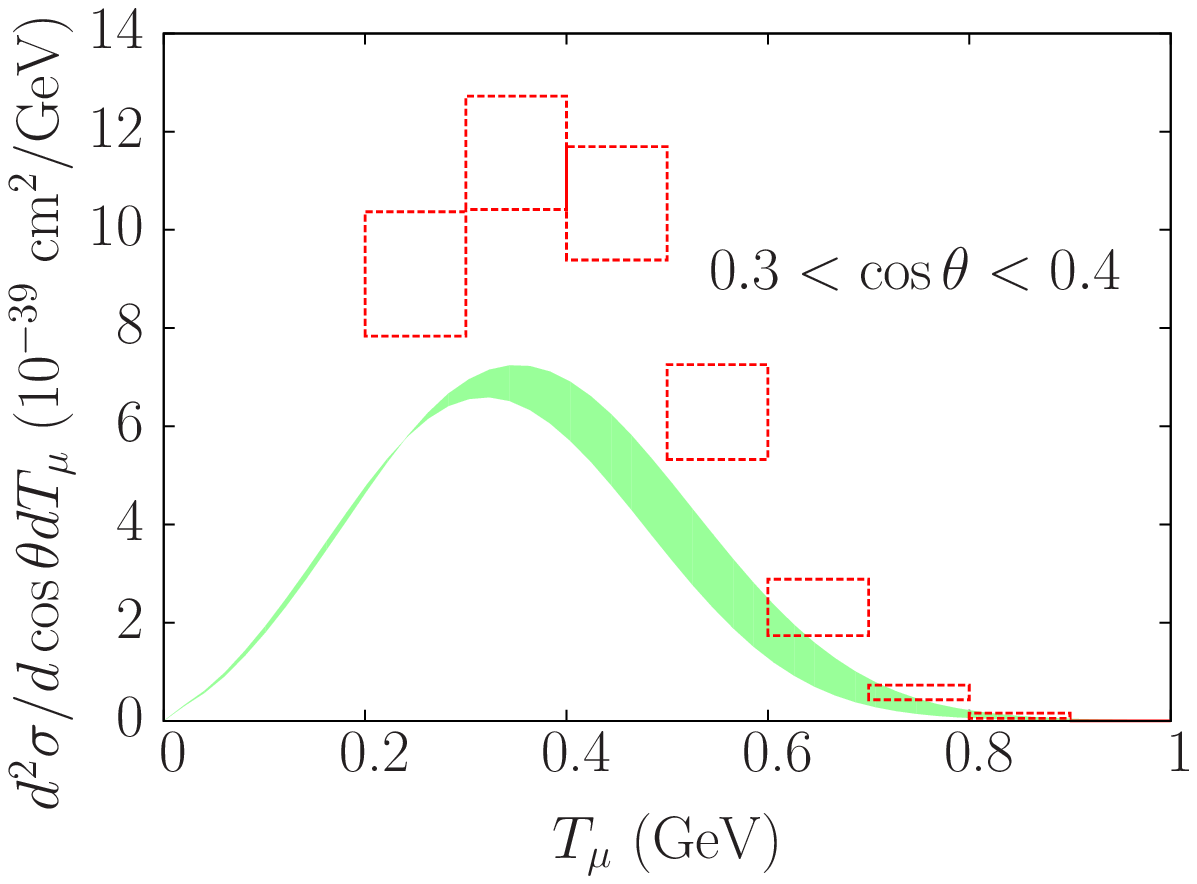}%
\includegraphics[scale=0.35]{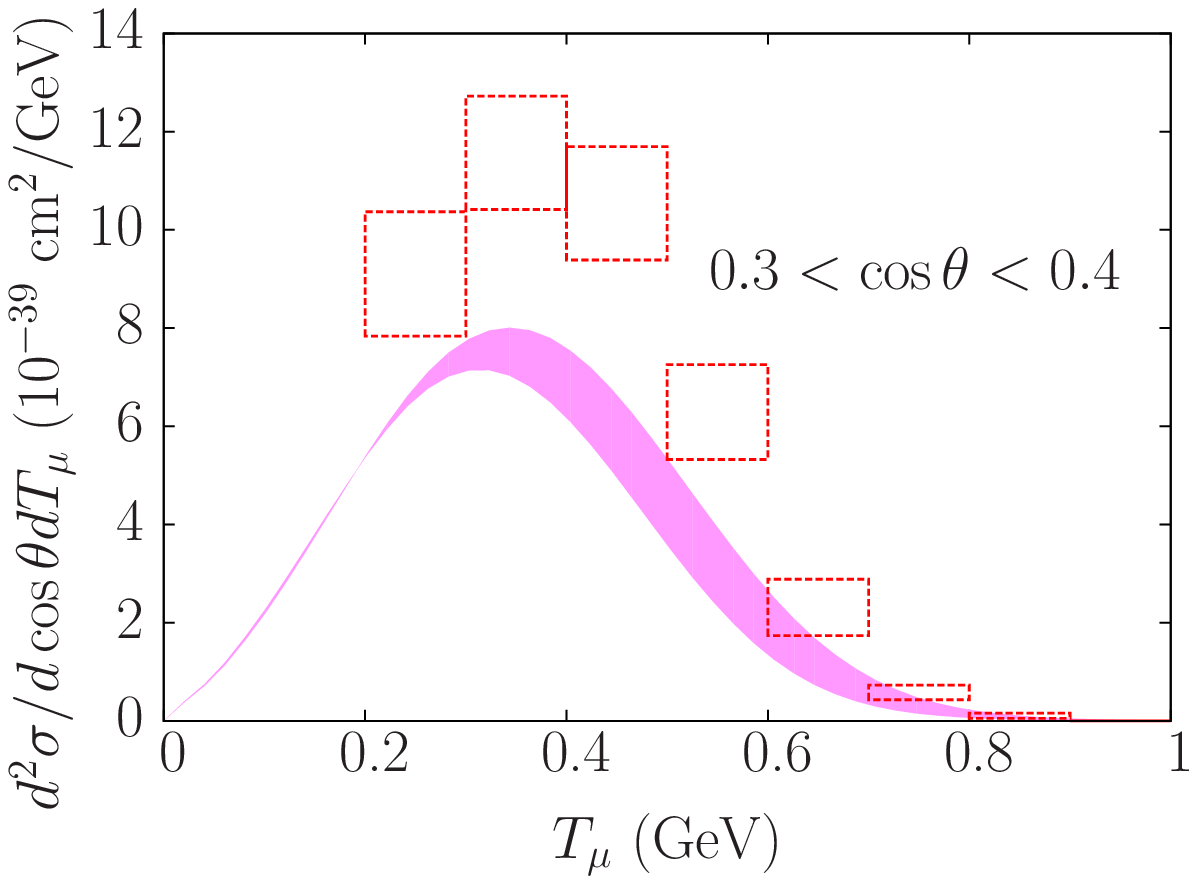}%
}}
\caption{Computed QE $(\nu_{\mu},\mu^-)$ cross section compared to Miniboone data. 
Left: SuSA. Right: SuSA + MEC}
\end{figure}

In Fig. 2 (right) we compare DEB+D and Woods-Saxon predictions for the
$^{12}$C$(\nu_\mu,\mu^-)$ differential QE cross section for 1 GeV
incident neutrinos and for several lepton scattering
angles. The SuSA reconstructed cross section is also shown, using
 the theoretical scaling function extracted from
$(e,e')$ results with the DEB+D model (curves in Fig. 2, left). 
For angles above 15$^{\rm o}$ the
SuSA is applicable and its predictions can be considered quite
reasonable, with small error.

The superscaling approach has been extended to inclusive neutrino 
scattering via the weak neutral current \cite{Ama06}. 
In the $(\nu,N)$ reaction
a nucleon is detected and the final neutrino must be integrated. 
We approximate the u-channel inclusive scattering cross section 
 \begin{equation}
\frac{d\sigma}{d\Omega_Ndp_N}
\simeq\overline{\sigma}^{(u)}_{sn}F^{(u)}(\psi^{(u)}).
\end{equation}
In Fig. 3 we see that the above factorization is a good approximation
in the RFG, and therefore one could extend the scaling analysis used
for CC reactions to NC scattering. In Fig. 3 we show examples of
SuSA predictions of NC cross sections using the phenomenological
$(e,e')$ scaling function.

Recently the effect of meson exchange currents (MEC) in $(e,e')$ for
high momentum transfer has been investigated \cite{Ama10a,Ama10b}, and
the two-particle emission (2p-2h) diagrams of Fig. 4 have been
extended to the CC weak interaction sector \cite{Ama10c}, in order to
explore the role of MEC in neutrino reactions.  In
Fig. 5 we show that inclusion of 2p-2h contributions yields results
for the QE $(\nu_{\mu},\mu^-)$ cross section
that are comparable with the recent MiniBooNE collaboration data
\cite{Agu10} for $\theta\leq 50^{\rm o}$, but lie below the data
at larger angles where the predicted cross sections are smaller. The
inclusion of the correlation diagrams which are required by gauge
invariance \cite{Ama10b} plus other relativistic effects might improve
the agreement with the data.

This work was
partially supported by DGI (Spain): FIS2008-01143,
FIS2008-04189, by the Junta de
Andaluc\'{\i}a, 
and part (TWD) by U.S.
Department of Energy under cooperative agreement DE-FC02-94ER40818.

\end{document}